\documentclass[12pt]{article}
\usepackage[a4paper, total={7in, 10in}]{geometry}
\usepackage[parfill]{parskip}
\usepackage{physics, tensor, float, subcaption}
\usepackage{graphicx}
\graphicspath{ {Plots/} }
\usepackage{jhep-mod}
\usepackage{bm}
\usepackage{soul}
\usepackage{amssymb,amsmath,amsthm}
\usepackage{mathrsfs}
\usepackage[utf8]{inputenc}
\usepackage{enumerate}
\usepackage{bigints}
\usepackage{xcolor}
\usepackage{appendix}
\usepackage{graphicx}
\usepackage{float}
\usepackage{tikz}
\usepackage{setspace}
\usepackage{cancel}
\usepackage{array}
\usepackage{tabulary}
\usepackage{doi}
\definecolor{purple}{rgb}{1,0,1}
\definecolor{lime}{HTML}{A6CE39} 

\newcommand{\blue}[1]{{\color{blue} #1}}

\definecolor{lime}{HTML}{A6CE39}
\newcommand{\orcidicon}{%
	\begin{tikzpicture}
	\draw[lime, fill=lime] (0,0) 
		circle [radius=0.16] 
		node[white] {{\fontfamily{qag}\selectfont \tiny ID}};
	\draw[white, fill=white] (-0.0625,0.095) 
		circle [radius=0.007];
	\end{tikzpicture}
	\hspace{-5mm}
}
\newcommand\orcidChris{{\href{https://orcid.org/0009-0000-3953-0461}{\orcidicon}}}
\newcommand\orcidMatt{{\href{https://orcid.org/0000-0003-1088-6485}{\orcidicon}}}

\newcommand{\be}{\begin{equation}}
\newcommand{\ee}{\end{equation}}

\def\sign{\mathrm{{sign}}}
\begin{document}
\newcommand{\arXiv}[1]{arXiv:\href{https://arxiv.org/abs/#1}{\color{blue}#1}}

\title{\vspace{-25pt}\huge{
\blue{Revisiting Schwarzschild's constant density star in isotropic coordinates}
}}


\author{\Large Christopher Simmonds\!\orcidChris\;}
\emailAdd{chris0simmonds@gmail.com}
\author{\!\!and \Large Matt Visser\!\orcidMatt$^{\dagger}$}
\emailAdd{matt.visser@sms.vuw.ac.nz}
\affiliation{School of Mathematics and Statistics, Victoria University of Wellington, \\
\null\qquad PO Box 600, Wellington 6140, New Zealand.}
\affiliation{$^\dagger$ Corresponding author.}
\renewcommand{\arXiv}[1]{arXiv:\href{https://arxiv.org/abs/#1}{\color{blue}#1}}
\def\L{{\mathcal{L}}}
\def\theta{\vartheta}
\def\phi{\varphi}

\abstract{ \\
Herein we shall revisit the venerable 110-year-old topic of Schwarzschild's constant density star, emphasizing that for many (though not quite all) purposes it is much easier to analyze this spacetime in isotropic coordinates (\emph{versus} the more usually adopted  Hilbert--Droste area coordinates). 
The relevant line element is particularly transparent, containing two simple rational functions of the radial coordinate, and the two physical parameters appearing in this line element are easily and readily interpretable in terms of the central density and central pressure of the star. 
Local properties in the stellar interior (such as the pressure profile) will be 
seen to be 
remarkably simple, though quasi-local properties like the Misner--Sharp mass are just a little bit  trickier.
Apart from its simplicity and clarity, the analysis is also of considerable pedagogical interest. For instance, there are a number of interesting special cases. 
Mathematically there is a perfectly good solution corresponding to a zero density star --- which can physically be interpreted as an explicit verification of the fact that pressure gravitates, even in the absence of mass-energy density. Additionally, there is a singular solution containing a naked singularity that satisfies all but one of the standard classical energy conditions. Furthermore you can even do both, combining zero density with a naked singularity --- so that pressure by itself can generate naked singularities --- at the cost of merely violating the dominant energy condition, the least physical of the standard energy conditions. We argue that many physically interesting features of Schwarzschild's star are very much under-appreciated.

\bigskip
\noindent
{\sc Date:}  Sunday 31 May 2026; \LaTeX-ed \today

 
\bigskip 
\bigskip
\hrule\hrule\hrule

}

\maketitle
\def\tr{{\mathrm{tr}}}
\def\diag{{\mathrm{diag}}}
\def\cof{{\mathrm{cof}}}
\def\pdet{{\mathrm{pdet}}}
\def\QED{ {\hfill$\Box$\hspace{-25pt}  }}
\def\d{{\mathrm{d}}}
\def\sign{\hbox{sign}}
\def\theta{\vartheta}
\def\phi{\varphi}

\parindent0pt
\parskip7pt

\clearpage
\null
\vspace{-75pt}
\section{Introduction}
The venerable subject of perfect fluid spheres in general relativity has an extensive 110 year 
history~\cite{Schwarzschild-star,Schwarzschild-star-translation,Tolman:1939, Wyman:1946, Wyman:1949, Buchdahl:1959, Adler:1974, Delgaty:1998, Rahman:2001, Lake:2002, Stephani:2003, Martin:2003, Boehmer:2005, Boonserm:2005a, Boonserm:2005b, Boonserm:2006, Boehmer:2007, Boonserm:2007a, Boonserm:2007b, Lake:2008, Loranger:2008, Grenon:2008, Griffiths:2009, Boonserm:2021, Arrechea:2023, Geodesy:2025}, dating back to Schwarzschild's constant density star~\cite{Schwarzschild-star}. Nevertheless, one still periodically encounters surprises.
Note that approximately 66\% of the known explicit solutions for perfect fluid spheres are given in terms of Hilbert--Droste area coordinates, while about 28\% are given in terms of isotropic coordinates, leaving only 6\% for other more unusual coordinates~\cite{Delgaty:1998}. 
In isotropic coordinates a static spherically symmetric line element takes the form
\begin{equation}
ds^2 = - A(r) \, dt^2 + B(r) \{ dr^2 + r^2[\d\theta^2+\sin^2\theta \; d\phi^2]\},
\end{equation}
whereas  in Hilbert--Droste area coordinates~\cite{Droste:1916, Droste:2002a, Droste:2002b, Hilbert:1916}
\begin{equation}
ds^2 = - A(\tilde r) \, dt^2 + B(\tilde r)  \, d\tilde r^2 + \tilde r^2[\d\theta^2+\sin^2\theta \; d\phi^2].
\end{equation}
(Schwarzschild himself used yet other multiple coordinate systems in deriving his vacuum solution, though the final expression he presented for the line element was functionally in Hilbert--Droste form~\cite{Schwarzschild:vacuum, Schwarzschild:vacuum-transation}.)
With a view to ultimately revisiting and updating the status of general perfect fluid spheres expressed in terms of  isotropic coordinates~\cite{Rahman:2001}, herein we shall focus on a much more specific topic: Schwarzschild's constant density star in isotropic coordinates~\cite{Wyman:1946}. 
 
 For many issues, especially purely local issues, it turns out that the use of isotropic coordinates greatly simplifies the discussion, with almost all results being expressible as simple rational polynomials.
 Quasi-local issues, such as the Misner--Sharp quasi-local mass, are a little trickier, but instructive.
 (Requiring nothing worse than square roots of rational polynomials.)
 Five special cases are of particular interest: the non-singular limit where the density is set to zero, a singular limit where the central pressure diverges,  a zero-density naked singularity, a star just on the edge of violating the dominant energy condition, and a sign-reversed variant of Einstein's static universe (sign-reversed Tolman I). We explore the role of the classical energy conditions in all of these situations. 
 
 In summary, Schwarzschild's constant density star expressed in isotropic coordinates is simple, 
 elegant, and pedagogically instructive.

\clearpage
\section{Schwarzschild's constant density star}

Consider the craftily chosen isotropic coordinate line element
\begin{equation}
ds^2 =
-{ (1+r^2/a^2)^2 
\over (1+r^2/b^2)^2} \; dt^2
+{ dr^2
+ r^2(d\theta^2+\sin^2\!\theta \;d\phi^2)
\over (1+r^2/b^2)^2}.
\end{equation}
Here the parameters $a$ and $b$ have the same dimensionality as $r$. 

\subsection{Local properties in the interior}

A straightforward calculation of the Einstein tensor shows 
\begin{equation}
G^t{}_t = -{12\over b^2};
\qquad
G^r{}_r = G^\theta{}_\theta = G^\phi{}_\phi = 
{4b^2(b^2 - 2 a^2) - 4 (2 b^2-a^2) r^2
\over  b^4(a^2+r^2)}.
\end{equation}
We note these are all simple (at worst quadratic) rational polynomials in $r$.

Invoking the Einstein equations, $G_{ab}=8\pi \, T_{ab}$ with $G_\mathrm{Newton}\to 1$, this represents a constant-density perfect fluid sphere with 
\begin{equation}
\rho(r) = \rho_c = {3 \over 2\pi b^2}; \qquad \qquad
p(r) 
= {b^2(b^2-2a^2)-(2b^2-a^2)r^2 \over 2\pi b^4(a^2+r^2)}.
\end{equation}
In terms of the central density $p_c=p(0)$ one has
\begin{equation}
p_c = {(b^2-2a^2) \over 2\pi a^2b^2};
\qquad
p(r) = p_c - {r^2(b^4-a^4)\over2 \pi a^2 b^4(a^2+r^2)}.
\end{equation}
To keep the density positive and finite one needs $b^2>0$. 
To keep the central pressure positive one additionally needs $b^2>2 a^2\geq 0$. 
To keep the central pressure finite one additionally needs $a^2> 0$. 

Under these conditions the pressure is monotone decreasing as one moves outwards, $p'(r) < 0$ is strictly negative:
\begin{equation}
p'(r) = -{r(b^4- a^4)\over\pi b^4 (a^2+r^2)^2}.
\end{equation}
This is still a rational polynomial, albeit now linear divided by quartic.

The surface of the star, defined by $p(r_s)=0$, is located at
\begin{equation}
r_s = \sqrt{b^2-2 a^2\over2b^2-a^2} \;\;  b < {b\over\sqrt{2}}.
\end{equation}
This square root occurring herein is the messiest part of the analysis.

\clearpage
All the way up to the surface of the fluid sphere, for $r\in[0,r_s)$, the Weyl tensor is identically zero, and up to symmetries the only non-zero components of the Riemann tensor are 
\begin{equation}
R^{r\theta}{}_{r\theta} = R^{r\phi}{}_{r\phi} = R^{\theta\phi}{}_{\theta\phi} = {4\over b^2};
\end{equation}
and
\begin{equation}
R^{tr}{}_{tr} = R^{t\theta}{}_{t\theta} = R^{t\phi}{}_{t\phi} = 
{2(a^2-b^2)(b^2-r^2)\over
b^4 (a^2+r^2)}.
\end{equation}
The Kretschmann scalar is
\begin{equation}
K = R_{abcd} \,R^{abcd} = R^{ab}{}_{cd} \, R^{cd}{}_{ab} = 
12 \left\{ {16 \over b^4} +{4 (a^2-b^2)^2 (b^2-r^2)^2
\over  b^8 (a^2+r^2)^2} \right\}.
\end{equation}
Note that (as expected for any static spacetime) the Kretschmann scalar is a (positive definite) sum of squares~\cite{Bronnikov:2013,Lobo:2020novel}. 
In view of this, for any static spacetime finiteness of the Kretschmann scalar  implies finiteness of all the mixed $R^{ab}{}_{cd}$ components of the Riemann tensor, which in turn implies regularity of the spacetime. Specifically for Schwarzschild's constant density star imposing both $b>0$ and $ (r^2+a^2)>0$ implies regularity.

All of these purely local properties are straightforward, involving rational polynomials at worst; 
the messiest feature is the square root occurring in the expression for the stellar radius $r_s$. 
In particular the pressure profile $p(r)$ in isotropic coordinates is considerably simpler than the corresponding result in terms of the more usual Hilbert--Droste area coordinates.

\bigskip

\subsection{Classical energy conditions}

The current status of the classical energy conditions is subtle. They are no longer, to current thinking, the absolute prohibitions they were often viewed as in the past. Instead they might best be viewed as diagnostics, warning signs, harbingers,  for the presence of unusual physics. For extensive general background regarding the energy conditions, see references~\cite{Parker:1973, Tipler:1978, Roman:1983, Borde:1987, Tsoubelis:1988, Yurtsever:1990, Ford:1994, Visser:1997a, Visser:1997b, Visser:1999, Barcelo:2002, Roman:2004, Fewster:2010, Martin-Moruno:2013a, Martin-Moruno:2013b, Curiel:2014, Martin-Moruno:2017, Lobo:2017, Lobo:2020, Kontou:2020, Hafemann:2025, Kontou:2026, Wang:2026a, Wang:2026b, Fliss:2026}. Note in particular that all of the standard energy conditions can easily be violated by quantum effects~\cite{GVP1,GVP2,GVP3,GVP4,GVP5}.

In the context of generic perfect fluids the classical energy conditions simplify to:
\begin{description}
\item[\qquad NEC:\qquad]  $\rho+p \geq 0$.
\item[\qquad WEC:\qquad]  \!$\rho+p \geq 0$ \quad and\quad  $\rho\geq 0$.
\item[\qquad SEC:\qquad]  \;$\rho+p \geq 0$ \quad and\quad $\rho+3p\geq 0$.
\item[\qquad DEC:\qquad]  $\rho\geq0$ \qquad\quad and\quad $|p|\leq \rho$.
\end{description}

\clearpage

In the current context --- since both the density and pressure are taken to be non-negative, this spacetime always satisfies the null, weak, and strong energy conditions, (NEC, WEC, and SEC).  
In counterpoint, the dominant energy condition (DEC), which is considerably less physically compelling~\cite{Barcelo:2002}, can sometimes be violated (in situations where $p>\rho$).  

To see this write
\begin{equation}
p(r) =  \rho_c +{b^2(b^2- 5a^2)-r^2(5b^2-a^2) \over2 \pi b^4 (a^2+r^2)}.
\end{equation}
The DEC is violated, $p(r)>\rho(r)$, once 
\begin{equation}
r < r_{\hbox{\tiny{DEC}}} = \sqrt{b^2-5a^2\over 5 b^2-a^2} \; b. 
\end{equation}
In order for the DEC violating region to be non-empty we need $b^2> 5 a^2$, a somewhat stronger condition than we needed impose to ensure positivity of the central pressure, ($b^2>2 a^2$). 
In particular, for $a^2\leq{1\over5} b^2$ there is no DEC violating region, the entire spacetime satisfies all of the standard classical energy conditions.

In counterpoint, recall
\begin{equation}
r_s = \sqrt{b^2-2a^2\over 2 b^2-a^2} \; b,
\end{equation}
whence for $\{a,b\}>0$ we have
\begin{equation}
0 \leq r_{\hbox{\tiny{DEC}}} < r_s  < {b\over\sqrt{2}}.
\end{equation}

Equivalently, we could just as well work in terms of the $w$-parameter
\begin{equation}
w(r)  = {p(r)\over\rho(r)} = {p_c\over\rho_c}- {2 r^2(b^4-a^4)\over3 a^2 b^2(a^2+r^2)}
=  {b^2-2a^2\over 3 a^2} - {r^2(b^4-a^4)\over3 a^2 b^2(a^2+r^2)}.
\end{equation}
The onset of DEC violations is now characterized by the condition $w(r)=1$, leading again to the same condition $r <   r_{\hbox{\tiny{DEC}}} $. 

In summary, under the conditions specified above, this spacetime always satisfies the null, weak, and strong energy conditions, (NEC, WEC, and SEC), but one can sometimes (sufficiently close to the centre) violate the (somewhat less physically compelling~\cite{Barcelo:2002}) dominant energy condition (DEC).

\clearpage
\subsection{Misner--Sharp quasilocal mass}
Observe that there are a few subtleties involved in adapting the Misner--Sharp quasi-local mass for use in isotropic coordinates.  Whereas the  Misner--Sharp quasi-local mass is commonly presented in an explicitly coordinate dependent manner, there is a natural covariant extension. 

\subsubsection{Prosaic methods}

The most prosaic and straightforward method of justifying the notion of Misner--Sharp quasi-local mass is simply to work in Hilbert--Droste area coordinates $(t,\tilde r,\theta,\phi)$ and write the line element as
\begin{equation}
ds^2 = - A(\tilde r) dt^2 + {d\tilde r^2\over 1-2m(\tilde r)/\tilde r} 
+ \tilde r^2 [\d\theta^2+\sin^2\theta\;d\phi]^2.
\end{equation}
Here $A(\tilde r)$ and $m(\tilde r)$ are, at this stage, just arbitrary functions of the Hilbert--Droste area coordinate $\tilde r$. 
It is then an easy exercise to verify that the Einstein equations imply
\begin{equation}
\rho(\tilde r) = {1\over4\pi \tilde r^2} \; {d m(\tilde r)\over d\tilde r}; 
\qquad\qquad
m(\tilde r) = \int_0^{\tilde r} 4\pi \bar r^2 \rho(\bar r) d\bar r,
\end{equation}
This can be rephrased as
\begin{equation}
g^{\tilde r\tilde r}\!(\tilde r) = 1-2m(\tilde r)/\tilde r; \qquad \qquad
m(\tilde r) = {\tilde r\over 2} \left[ 1- g^{\tilde r\tilde r}(\tilde r) \right].
\end{equation}
Misner and Sharp~\cite{Misner:1964} then realized that this last formula can be made fully covariant, and can be re-expressed in arbitrary spherically symmetric coordinates. (See also the slightly later discussion by Misner and Hernandez~\cite{Hernandez:1966}.)

\subsubsection{Invariant methods}

Consider the most general spherically symmetric block-diagonal line element. (Not even necessarily static.) 
Let us work in $x^a = (x^i;\theta,\phi)$ coordinates. We have:
\begin{equation}
ds^2 = \gamma_{ij} \,dx^i\,  dx^j + \tilde r(x)^2 [ d\theta^2+ \sin^2\theta \; d\phi^2].
\end{equation}
Here $\tilde r(x)$ is the so-called ``area radius'', the covariant extension of the Hilbert--Droste area coordinate.
The 2-spheres of spherical symmetry have area $A(x)=4\pi \, \tilde r(x)^2$. 
Then the general formula for the Misner--Sharp quasilocal mass is~\cite{Misner:1964} 
\begin{equation}
m(\tilde r) = {\tilde r\over2} \;\left[1- \gamma^{ij} \;\nabla_i \tilde r \; \nabla_j \tilde r \right];
\qquad \hbox{(general)}.
\end{equation}

Now consider a slightly more specific  static and diagonal line element
\begin{equation}
ds^2 = g_{tt}(r) dt^2 + g_{rr}(r) dr^2 + g_{\theta\theta}(r)\left[ d\theta^2+ \sin^2\theta \; d\phi^2\right].
\end{equation}
Then the general formula for the Misner--Sharp quasilocal mass reduces to
\begin{equation}
m(r) = {\sqrt{g_{\theta\theta}(r)}\over 2}\left[ 1 - g^{rr}(r) \;\partial_r \sqrt{g_{\theta\theta}(r)} \; \partial_r \sqrt{g_{\theta\theta}(r)} \right];
\qquad \hbox{(diagonal static)}.
\end{equation}
That is
\begin{equation}
m(r) = {\sqrt{g_{\theta\theta}(r)}\over 2}
\left[ 1 - {1\over4} \;{[\partial_r g_{\theta\theta}(r)]^2\over 
g_{rr}(r) \, g_{\theta\theta}(r)}  \right];
\qquad \hbox{(diagonal static)}.
\end{equation}
In Hilbert--Droste curvature coordinates $g_{\theta\theta}(r)\to r^2$ and we have the usual
\begin{equation}
m(r) = {r\over 2}
\left[ 1 -  g^{rr}(r) \right]; \qquad 
g_{rr}(r) = {1\over 1 - 2m(r)/r}; \qquad \hbox{(Hilbert--Droste)}.
\end{equation}
In isotropic coordinates $g_{\theta\theta}(r)\to r^2 g_{rr}(r)$ and we have the perhaps unexpected
\begin{equation}
m(r) = {r\over 2}\, \sqrt{g_{rr}}
\left[ 1 -  \left( 1 + {1\over2}\,{\partial_r g_{rr}\over g_{rr}} \right)^2 \right];  \qquad \hbox{(isotropic)}.
\end{equation}
That is
\begin{equation}
m(r) = -{r\over 2}\, \sqrt{g_{rr}}
\left[{\partial_r g_{rr}\over g_{rr}} + {1\over 4} \left({\partial_r g_{rr}\over g_{rr}} \right)^2 \right];  \qquad \hbox{(isotropic)}.
\end{equation}
While the explicit form of this isotropic coordinate version of the Misner--Sharp mass is perhaps a little unexpected, it can easily be sanity checked by considering isotropic version of Schwarzchild's vacuum solution:
\begin{equation}
ds^2 = 
-\left(1 -{m\over2r}\over1+{m\over2r}\right)^2 dt^2 +
 \left(1+{m\over 2r}\right)^4 [ dr^2+ r^2(d\theta^2+\sin^2\theta \,d\phi^2)].
\end{equation}
In this situation $g_{rr}(r) \to \left(1+{m\over 2r}\right)^4$ and, as it should,  we see $m(r)\to m$.

For the isotropic coordinate version of Schwarzschild's constant density star we have already seen $g_{rr}(r) \to (1+r^2/b^2)^{-2}$,  and so we now see that
\begin{equation}
m(r) = {2 r^3\over b^2(1+r^2/b^2)^3}.
\end{equation}
At least this final expression for $m(r)$ is relatively simple.

\clearpage
This expression is also a lot less mysterious if one expresses it directly in terms of the area radius $\tilde r(r)$, since in the current context 
\begin{equation}
\tilde r(r) = \sqrt{g_{\theta\theta}(r) }=  r\sqrt{g_{rr}(r) } = {r \over 1+ r^2/b^2}.
\end{equation}
Thence 
\begin{equation}
m(\tilde r(r)) = {2\, [\tilde r(r)]^3\over b^2}= {4\pi\over3} \,\rho_c \; [\tilde r(r)]^3.
\end{equation}

Perhaps unfortunately, at the surface of the star one finds
\begin{equation}
\tilde r_s = \tilde r(r_s) = {b\over 3} \left( \sqrt{2b^2-a^2} \, \sqrt{b^2-2 a^2}  \over b^2-a^2 \right),
\end{equation}
so that
\begin{equation}
m_s = m(\tilde r(r_s)) = m(\tilde r_s) = {2 \, (2b^2-a^2)^{3/2} \, (b^2-2 a^2)^{3/2} b \over 27 (b^2-a^2)^3}.
\end{equation}
That is
\begin{equation}
m_s = m(\tilde r_s) = {2 b\over27}  \left( \sqrt{2b^2-a^2} \, \sqrt{b^2-2 a^2}  \over b^2-a^2 \right)^3.
\end{equation}
The only reason for this slightly messy formula are the square roots appearing in the expressions for $r_s$ and $\tilde r_s$.
A somewhat less messy result, eliminating the square roots,  is to note
\begin{equation}
{2m_s\over\tilde r_s} = {4\over 9} \;\;{(2b^2-a^2) \, (b^2-2 a^2)  \over (b^2-a^2)^2 }
={8\over 9}- {a^2 b^2\over  (b^2-a^2)^2 } \leq {8\over9}.
\end{equation}
The numerical constant ${8\over9}$ that shows up here is a standard result --- it is a specific example of the famous Buchdahl--Bondi bound~\cite{Buchdahl, Bondi, Arrechea:2021, Arrechea:2024, Bueno:2025, Garcia-Moreno:2025, Cattoen:2005}.

\subsection{Parameterizing the line element in terms of $p_c$ and $\rho_c$}

It is sometimes useful to parameterize the line element in terms of the central pressure and density, or central density and the central $w$ parameter. From the discussion above we see
\begin{equation}
{1\over b^2} = {2\pi \rho_c\over 3}; \qquad 
{1\over a^2} = {2\pi (2\rho_c+3p_c)\over 3}=
{2\pi \rho_c\over 3}(2+3 w_c);
\end{equation}
Then the line element can be recast as
\begin{equation}
ds^2 =
-{ (1+{2\pi\over3} (2\rho_c+3p_c) \,r^2)^2 
\over (1+{2\pi\over3}  \rho_c \,r^2)^2} dt^2
+{ dr^2
+ r^2(d\theta^2+\sin^2\theta \,d\phi^2)
\over (1+{2\pi\over3}  \rho_c \,r^2)^2}.
\end{equation}
Equivalently
\begin{equation}
ds^2 =
-{ (1+{2\pi\over3} \rho_c (2+3w_c) \,r^2)^2 
\over (1+{2\pi\over3}  \rho_c \,r^2)^2} dt^2
+{ dr^2
+ r^2(d\theta^2+\sin^2\theta \,d\phi^2)
\over (1+{2\pi\over3}  \rho_c \,r^2)^2}.
\end{equation}

Then
\begin{equation}
p(r) 
= {p_c - {2\pi\over3} \rho_c(\rho_c+2p_c) r^2\over1 + 2\pi(p_c+{2\over3}\rho_c) r^2}
= \rho_c \; {w_c - {2\pi\over3} \rho_c(1+2w_c) r^2\over1 + {2\pi\over3}\rho_c(2+3 w_c) r^2}.
\end{equation}
That is
\begin{equation}
p(r) 
= p_c \; {1 - {2\pi\over3} \rho_c (2+w_c^{-1}) r^2\over1 + {2\pi\over3}\rho_c(2+3 w_c) r^2}
= p_c \; {1 - {2\pi\over3} p_c w_c^{-1}(2+w_c^{-1}) r^2\over1 + {2\pi\over3}p_c w_c^{-1}(2+3 w_c) r^2}.
\end{equation}

In this parameterization the surface of the star is at
\begin{equation}
r_s^2 = {p_c\over {2\pi\over3} \rho_c(\rho_c+2p_c) }
= {3 w_c\over 2\pi \rho_c (1+2w_c)} < { 3\over 4\pi \rho_c }.
\end{equation}
(Where the bound assumes $w_c > 0$.)

In this parameterization the onset of DEC violation occurs once
\begin{equation}
r^2 < r_{\hbox{\tiny{DEC}}}^2 = 
{3 (p_c-\rho_c)\over {2\pi} \rho_c 
(5 p_c+3 \rho_c) }
= {3 (w_c-1)\over 2\pi \rho_c (5w_c+3)} 
< { 3/5\over 2\pi \rho_c }.
\end{equation}

The quasilocal mass $m(r)$ can be written in the extremely suggestive form
\begin{equation}
m(r) = {{4\pi\over 3} \rho_c r^3 \over (1 +{2\pi\over 3} \rho_c r^2)^3},
\end{equation}
where it is very clearly a relativistic ``enhancement'' of the naive Newtonian result $m_{Newton}(r) = {4 \pi\over 3} \rho_c r^3$.

Indeed
\begin{equation}
m(r) = {m_{Newton}(r)  \over \left(1 +{m_{Newton}(r) \over 2r}\right)^3}.
\end{equation}

At the surface of the star
\begin{equation}
m_s = m(r_s) = \sqrt{6\over \pi \rho_c} \left( \sqrt{w_c(1+2 w_c)} \over 1+3 w_c\right)^3
\leq {4\over 27} \sqrt{3\over \pi \rho_c}.
\end{equation}

Similarly, from
\begin{equation}
\tilde r(r) = {r \over 1+{2\pi\over3}  \rho_c \,r^2},
\end{equation}
we have
\begin{equation}
\tilde r_s = \tilde r(r_s) = {r_s \over 1+{3w_c\over1+2w_c}}= \sqrt{3\over2\pi\rho_c} \left( \sqrt{w_c(1+2 w_c)} \over 1+3 w_c\right),
\end{equation}
and so
\begin{equation}
{2m_s\over\tilde r_s} = {8w_c^2+4w_c\over(3w_c+1)^2}
= {8\over 9} - {4\over 9} \,{(3w_c+2)\over(3w_c+1)^2} < {8\over9}.
\end{equation}
We again see the occurrence of the Buchdahl--Bondi bound~\cite{Buchdahl,Bondi, Arrechea:2021, Arrechea:2024, Bueno:2025, Garcia-Moreno:2025,Cattoen:2005}, now arising in the limit $w_c\to\infty$.

\subsection{Junction conditions at the surface}

Junction conditions at the surface of the star are commonly phrased in terms of the Israel--Lanczos--Sen formalism~\cite{Israel,Lanczos,Sen}. (See also~\cite{Barcelo:2000-moduli, Das:2003, Lobo:2003, Lobo:2004, Marolf:2005, Chu:2021, Lake:2017,  Senovilla:2026, Lobo:2025}.) For practical purposes we provide a condensed summary:
\begin{itemize}
\item 
Keep the pressure continuous (zero) at the surface (since otherwise there are infinite pressure gradients that would blow the surface off). 
\item 
Keep the area coordinate $\tilde r(r)$  continuous at the surface (since otherwise the areas of the 2-spheres of spherical symmetry become discontinuous).  Note there is no particular need for the isotropic $r$ coordinate itself to be continuous, that's just a coordinate, a label; instead it is $g_{\theta\theta}(r)$ that must be continuous.
\item 
Keep the quasilocal mass $m(r)$ continuous (since otherwise one is implicitly asserting the presence of   an [infinitely] thin layer of matter at the surface). 
\end{itemize}
While the discussion could be rephrased in terms of continuity of the first and second fundamental forms (intrinsic and extrinsic curvatures) we find the version above to be more physically compelling.

\subsection{Speed of sound}
It is often claimed that Schwarzschild's constant density star implies an infinite speed of sound. 
This is actually a misapprehension based on unwarranted implicit assumptions.
The usual  argument
\begin{equation}
c_{sound}^2 = {dp\over d\rho} \to {dp/dr\over d\rho/dr} \to \infty,
\end{equation}
relies on an unstated assumption that one is dealing with a highly mixed fluid star described by a single global equation of state $p(\rho)$. 
But in Schwarzschild's constant density star no such global equation of state can possibly exist.

The best interpretation one can come up with for Schwarzschild's constant density star is that it is \emph{stratified}, 
with an equation of state of the form $p(\rho,\mu)$ where $\mu(r)$ stands for some physical variable characterizing the composition of the fluid (chemical potential, temperature, entropy density, number density, \emph{etcetera}). Then a correct statement is 
\begin{equation}
c_{sound}^2 = \left.{\partial p\over \partial\rho}\right|_\mu,
\end{equation}
which can be adjusted as desired and certainly need not be infinite. 

Specifically, rewriting the equation of state as $\rho(p,\mu)$, we see
\begin{equation}
0 = {d\rho\over dr} = 
\left.{\partial \rho\over \partial p}\right|_\mu {dp\over dr} + 
 \left.{\partial \rho\over \partial \mu}\right|_p {d\mu\over dr}=
 c_{sound}^{-2} {dp\over dr} + 
 \left.{\partial \rho\over \partial \mu}\right|_p {d\mu\over dr}.
\end{equation}
Thence, rearranging
\begin{equation}
c_{sound}^{2} = \left.{\partial p\over \partial\rho}\right|_\mu= 
-  \left.{\partial \mu\over \partial \rho}\right|_p  \;  {dp/ dr\over d\mu /dr},
\end{equation}
which can be adjusted as desired and certainly need not be infinite. (The minus sign here is correct and is a consequence of the fact that this last equality is an example of a thermodynamic Maxwell relation.)
In short, the often repeated claims that Schwarzschild's constant density star implies an infinite speed of sound, are on closer inspection, utterly unfounded.

\clearpage
\section{Various special cases}
One of the very nice features of the isotropic coordinate version of Schwarzschild's constant density star is that it becomes rather straightforward to analyze various pedagogically interesting special cases. 

\subsection{Schwarzschild's zero density star (pressure gravitates)}
The zero density limit corresponds to $b \to\infty$, in which case the line element is
\begin{equation}
ds^2 =
 -(1 +r^2/a^2)^2 dt^2 + dr^2 + r^2(d\theta^2+\sin^2\theta \,d\phi^2),
\end{equation}
and we have
\begin{equation}
\rho = 0; \qquad 
p(r) = {1\over 2\pi (a^2+r^2)}.
\end{equation}
This spacetime everywhere satisfies the null, weak, and strong energy conditions, 
however it now everywhere violates the, (somewhat less physically compelling~\cite{Barcelo:2002}), dominant energy condition.
Pedagogically this is interesting in that it explicitly verifies that pure pressure gravitates, even in the absence of density. 
Note that the central pressure is still finite and we can write
\begin{equation}
p_c={1\over2\pi a^2}; \qquad p(r) = {p_c\over1+ 2\pi \,p_c \, r^2}.
\end{equation}

Note that 
the ``star'' no longer has a surface and extends all the way out to spatial infinity.
The Weyl tensor is now identically zero throughout the entire spacetime, and up to symmetries the only non-zero components of the Riemann tensor are 
\begin{equation}
R^{tr}{}_{tr} = R^{t\theta}{}_{t\theta} = R^{t\phi}{}_{t\phi} = 
-{2\over (a^2+r^2)}.
\end{equation}
The Kretschmann scalar is
\begin{equation}
K = R_{abcd} \; R^{abcd} = +{48 \over (a^2+r^2)^2}.
\end{equation}
Because one now has $g_{rr}\equiv 1$, the Misner--Sharp quasilocal mass is identically zero, $m(r)\equiv 0$.

\subsection{Schwarzschild's singular star}
The singular limit corresponds to first rescaling $t\to at$, and then setting $a\to 0$, while keeping $b$ finite, 
to obtain the line-element
\begin{equation}
ds^2 =
-{ r^4 \over (1+r^2/b^2)^2} dt^2
+{ dr^2
+ r^2(d\theta^2+\sin^2\theta \,d\phi^2)
\over (1+r^2/b^2)^2}.
\end{equation}
A straightforward calculation of the Einstein tensor now shows 
\begin{equation}
G^t{}_t = -{12\over b^2};
\qquad
G^r{}_r = G^\theta{}_\theta = G^\phi{}_\phi = 
{4(b^2 - 2r^2) \over b^2r^2}.
\end{equation}
This represents a perfect fluid sphere with 
\begin{equation}
\rho(r) = \rho_c = {3 \over 2\pi b^2}; \qquad 
p(r) 
= {b^2-2r^2 \over 2\pi  b^2r^2} 
= {1\over2\pi}\left({1\over r^2}-{2\over b^2}\right).
\end{equation}
The surface is at $r_s=b/\sqrt2$ and
\begin{equation}
w(r) = {p(r)\over \rho(r)} = {b^2\over 3 r^2}-{2\over3} \;\in\,(0,\infty).
\end{equation}
DEC violations occur once $r<  r_{\hbox{\tiny{DEC}}} = b/\sqrt{5}$, but all of the other usual energy conditions (NEC, WEC, SEC) are satisfied throughout the star.

The Riemann components are
\begin{equation}
R^{r\theta}{}_{r\theta} = R^{r\phi}{}_{r\phi} = R^{\theta\phi}{}_{\theta\phi} = {4\over b^2};
\end{equation}
and
\begin{equation}
R^{tr}{}_{tr} = R^{t\theta}{}_{t\theta} = R^{t\phi}{}_{t\phi} = 
{2(r^2-b^2)\over
b^2 r^2}.
\end{equation}
The Kretschmann scalar is
\begin{equation}
K = R_{abcd} R^{abcd} = {192 \over b^4}
+{48 (b^2-r^2)^2  \over b^4 r^4}.
\end{equation}
The quasilocal mass remains nonzero
\begin{equation}
m(r) = {2 r^3\over b^2(1+r^2/b^2)^3}.
\end{equation}
Note
\begin{equation}
m_s=m(r_s) \to {\sqrt{2}\over 3} b,
\end{equation}
while
\begin{equation}
\tilde r_s=\tilde r(r_s) \to{4\sqrt{2}\over 27} b,
\end{equation}
and so
\begin{equation}
{2m_s\over\tilde r_s}\to {8\over9}.
\end{equation}
The numerical constant ${8\over9}$ is a standard result --- it is again a specific instance of the famous Buchdahl--Bondi bound~\cite{Buchdahl, Bondi, Arrechea:2021, Arrechea:2024, Bueno:2025, Garcia-Moreno:2025,Cattoen:2005}.

\clearpage

\subsection{Schwarzschild's zero density singular star}
As regards the line-element the singular limit corresponds to first rescaling $t\to at$, and then setting $a\to 0$, while simultaneously demanding $b\to \infty$,  to obtain
\begin{equation}
ds^2 =
- r^4 dt^2+ dr^2 + r^2(d\theta^2+\sin^2\theta \,d\phi^2).
\end{equation}
A straightforward calculation of the Einstein tensor now shows 
\begin{equation}
G^t{}_t = 0;
\qquad
G^r{}_r = G^\theta{}_\theta = G^\phi{}_\phi = 
{4 \over r^2}.
\end{equation}
This represents a perfect fluid sphere with 
\begin{equation}
\rho(r) = 0; \qquad  p(r)  = {1\over 2\pi r^2}.
\end{equation}
There is no surface, $r_s\to \infty$, and the only non-zero Riemann components are
\begin{equation}
R^{tr}{}_{tr} = R^{t\theta}{}_{t\theta} = R^{t\phi}{}_{t\phi} = 
-{2\over r^2}.
\end{equation}
The Kretschmann scalar is
\begin{equation}
K = {48 \over r^4}.
\end{equation}
NEC, WEC, and SEC are satisfied --- the only one of the standard energy conditions that is violated is the DEC.

\subsection{Schwarzschild's conservative extremal star}
Another pedagogically useful example is to ask for the most extremal constant density star that still satisfies \emph{all} of the standard classical energy conditions --- including the DEC. From the discussion regarding energy conditions above, this happens when $a^2={1\over5} b^2$.
The line element becomes
\begin{equation}
ds^2 =
-{ (1+5 r^2/b^2)^2 
\over (1+r^2/b^2)^2} \; dt^2
+{ dr^2
+ r^2(d\theta^2+\sin^2\!\theta \;d\phi^2)
\over (1+r^2/b^2)^2}.
\end{equation}
Then
\begin{equation}
\rho_c=p_c= {3\over 2\pi b^2}; \qquad 
p(r) = p_c \;{1-3r^2/b^2\over1+5r^2/b^2} = p_c\;{1-2\pi p_c r^2\over 1 +{5\over3} 2\pi p_c r^2}.
\end{equation}
The surface of the star is now at $r_s=b/\sqrt3= 1/\sqrt{2\pi p_c}$, and 
up to symmetries the only non-zero components of the Riemann tensor are 
\begin{equation}
R^{r\theta}{}_{r\theta} = R^{r\phi}{}_{r\phi} = R^{\theta\phi}{}_{\theta\phi} = {4\over b^2};
\end{equation}
and
\begin{equation}
R^{tr}{}_{tr} = R^{t\theta}{}_{t\theta} = R^{t\phi}{}_{t\phi} = 
-{8(b^2-r^2)\over
b^2 (b^2+5r^2)}.
\end{equation}
The Kretschmann scalar is
\begin{equation}
K = R_{abcd} \,R^{abcd} = R^{ab}{}_{cd} \, R^{cd}{}_{ab} = 
12 \left\{ {16 \over b^4} +{64 (b^2-r^2)^2
\over  b^4 (b^2+5r^2)^2} \right\},
\end{equation}
and is again manifestly a sum of squares.

This example satisfies all energy conditions, (though it is on the verge of violating the DEC), and is particularly elegant and straightforward.

\subsection{(Sign reversed) Einstein static universe/Tolman I solution}
The special case $a=b$ corresponds to the line element
\begin{equation}
ds^2 =- dt^2
+{ dr^2+ r^2(d\theta^2+\sin^2\!\theta \;d\phi^2)
\over (1+r^2/b^2)^2}.
\end{equation}
The Einstein tensor is 
\begin{equation}
G^t{}_t = -{12\over b^2};
\qquad
G^r{}_r = G^\theta{}_\theta = G^\phi{}_\phi = -{4\over b^2}.
\end{equation}
This represents a perfect fluid sphere with 
\begin{equation}
\rho(r) = \rho_c = {3 \over 2\pi b^2}; \qquad 
p(r) = p_c = -{1\over 2\pi b^2}= -{\rho_c\over 3}; \qquad 
\rho_c + 3 p_c =0.
\end{equation}
Note that, since $\rho_c + 3 p_c =0$, this particular spacetime is right on the boundary of violating the SEC.

\enlargethispage{20pt}
This is a sign-reversed version of the Einstein static universe/Tolman I solution~\cite{Tolman:1939}.
As written above, the density is positive, the pressure is negative, and the spatial slices are hyperbolic 3-planes. 

The traditional version of the Einstein static universe/Tolman I solution corresponds to the additional formal substitution $b^2 \to - b^2$,  so that 
\begin{equation}
ds^2 =- dt^2
+{ dr^2+ r^2(d\theta^2+\sin^2\!\theta \;d\phi^2)
\over (1-r^2/b^2)^2}.
\end{equation}
With the sings as chosen above NEC and WEC are satisfied, SEC is marginally satisfied, and DEC is satisfied. 
In this traditional version, the density is negative, the pressure is positive, and the spatial slices are 3-spheres with $r\in(0,b)$. 

\section{Conclusions}

In this article we have revisited Schwarzschild's well-known 110-year-old  constant density star~\cite{Schwarzschild-star, Schwarzschild-star-translation}, specifically working in isotropic coordinates~\cite{Wyman:1946}. 
The reasons for doing this are two-fold: (1) using isotropic coordinates makes the discussion a lot simpler, and, as we have seen above, is pedagogically interesting for a number of reasons; (2) with a view to the future, we very strongly suspect that the use of isotropic coordinates  for generic perfect fluid spheres may also prove advantageous. Work along these lines is ongoing. 


\bigskip
\bigskip
\hrule\hrule\hrule

\clearpage
\hrule\hrule\hrule
\addtocontents{toc}{\bigskip\hrule\hrule\hrule}

\vspace{-25pt}
\setcounter{secnumdepth}{0}
\section[\hspace{14pt}  References]{}
%

\end{document}